\documentclass[twocolumn,showpacs,preprintnumbers,amsmath,amssymb]{revtex4}
\usepackage{graphicx}
\usepackage{dcolumn}
\usepackage{bm}
\begin{document}
\def\be{\begin{equation}}
\def\ee{\end{equation}}
\title{On the local Lorentz invariance in $N=1$ supergravity}

\author{Alfredo Mac\'{\i}as}
 \email{amac@xanum.uam.mx}
\affiliation{Departamento de F\'{\i}sica, Universidad Autonoma Metropolitana--Iztapalapa,\\
A.P. 55--534, M\'exico D.F. 09340, M\'exico}
\author{Hernando Quevedo}
 \email{quevedo@nuclecu.unam.mx}
\affiliation{Instituto de Ciencias Nucleares,
Universidad Nacional Aut\'onoma de M\'exico\\
A.P. 70-543,  M\'exico D.F. 04510, M\'exico}
\author{Alberto S\'anchez}
 \email{asan@xanum.uam.mx}
 \affiliation{Departamento de F\'{\i}sica, Universidad Autonoma Metropolitana--Iztapalapa,\\
A. P. 55--534, M\'exico D.F. 09340, M\'exico}

\date{\today}

\begin{abstract}

We discuss the local Lorentz invariance in the
context of $N=1$ supergravity and show that a previous 
attempt to find explicit solutions to the Lorentz constraint
in terms of $\gamma-$matrices is not correct. 
We improve that solution by using a different representation
of the Lorentz operators in terms of the generators 
of the rotation group, and show its compatibility with
the matrix representation of the fermionic field. We find 
the most general wave functional that satisfies the 
Lorentz constraint in this representation.

{\bf File: lor.tex; 05.05.2005}
\end{abstract}


\maketitle


\section{Introduction}
\label{sec1}
In the Lagrangian for  $d=4$,   $N=1$ supergravity
\begin{equation}
\label{lag1} {\cal{L}}=\frac{1}{2} \sqrt{-g} R- \frac{i}{2}
\epsilon^{\lambda\mu\nu\rho} \bar{\psi}_\lambda \gamma_{5}
\gamma_{\mu} D_{\nu} \psi_{\rho}\, ,
\ee
\be
D_{\nu}=\partial_\nu +\frac{1}{2} \omega_{\nu A B}
\sigma^{AB}\, ,
\end{equation}
the fields are the local tetrad ${\bf e}$
and the Rarita-Schwinger gravitino field ${\bf \psi}$
(here we follow closely the conventions and notations of \cite{msq}).
Here $\omega_{\nu AB}$ is the connection associated with the
local vierbein ${\bf e}$ and 
\be
\sigma^{AB}=\frac{1}{4}(\gamma^A \gamma^B-\gamma^B \gamma^A)
\label{lorg}
\ee
are the Lorentz generators in terms of the standard $\gamma-$matrices.
The corresponding canonical $3+1$ decomposition of spacetime
leads to a constrained system determined by the Hamiltonian
\begin{equation} 
H= e^A{}_0\, {\cal H}_A
+ \frac{1}{2} \omega_{0}{}^{AB}\,{\cal J}_{AB}
+ {\overline \psi}_0\, {\cal S} \label{sgham}\, ,
\end{equation}
to which one can apply 
Dirac's canonical quantization procedure \cite{Dirac65}. 
As a result one obtains a system of three constraint
operators which must annihilate the ground state of the wave functional \cite{pi78}. The 
constraint operators are associated with the gauge symmetries
of the system, namely spacetime diffeomorphisms $({\cal H}_A)$, 
local Lorentz invariance $({\cal J}_{AB})$, and supersymmetry $({\cal S})$. 
Accordingly, the constraints are   
\be
{\cal H}_A \Psi[{\bf e}, {\bf \psi}] = 0, \quad  
{\cal J}_{AB}\Psi[{\bf e}, {\bf \psi}] = 0, \quad
{\cal S}\Psi[{\bf e}, {\bf \psi}] = 0 \ ,
\ee
where $\Psi[{\bf e}, {\bf \psi}]$ represents the wave functionals
(wave function of the universe) which depend in general on the 
bosonic ${\bf e}$ and fermionic ${\bf \psi}$ fields.  

It is well known that the constraint operators satisfy Teitelboim's
algebra \cite{tei77} (we quote only the for us relevant term)
\be
\{ {\cal S}(x), {\overline {\cal S}} (x^\prime)\} = \gamma^A{\cal H}_A
\delta(x,x^\prime) \ ,
\ee
from which it follows that the constraint 
${\cal H}_A \Psi[{\bf e}, {\bf \psi}] =0$ is satisfied identically 
once the supersymmetric constraint ${\cal S}
\Psi[{\bf e}, {\bf \psi}]=0$ is fulfilled. 
Consequently, only the Lorentz and supersymmetric constraints
need to be considered in order to obtain wave functionals.

In this work we will focus on the Lorentz constraint. 
The technical difficulties in the manipulation of this constraint are
drastically reduced due to the fact that in an appropriate basis (see 
below) it does not depend explicitly 
on the bosonic variables. 
In principle, it is very easy to satisfy the Lorentz constraint
 since it simply states that the wave functional 
$\Psi[{\bf e}, {\bf \psi}]$  must be a Lorentz invariant,
a condition that can be satisfied, for instance, by demanding
that the wave functional contains the bosonic ${\bf e}$
and fermionic   ${\bf \psi}$ variables only in local Lorentz invariant 
combinations \cite{death84}. In particular, one can take the 
wave functional as an expansion in terms of scalars constructed 
by means of bosonic and fermionic variables. 
Since the constraints of $N=1$ supergravity are homogeneous in the
gravitino field ${\psi}$, one can search for solutions containing 
homogeneous functionals of order $\psi^n$. These states are called
Grasmann number $n$ states. In this case, it
is necessary to consider that there are no physical states that 
are purely bosonic or have fixed Grassmann number $n$ \cite{cfop}.
In fact, any physical state in $N=1$ supergravity 
must contain fermionic degrees of freedom and must have infinite
Grassmann number $n$. Notice that all this is valid only for the
ground state of the wave functional. In order to analyze  excited 
states, one needs to explore the Pauli-Lubanski operator and its
eigenvalues. This would allow to investigate solutions with spin and/or
helicity \cite{kaku}.

On the other hand, it has been claimed \cite{mms98,or98} 
that the Lorentz constraint needs to be solved
explicitly in completely general terms, 
without making any assumptions on the Lorentz
behavior of wave functional. 
In \cite{mms98,mmsotros}, different minisuperspace Bianchi models
were analyzed, finding explicit solutions to the Lorentz
constraint by using a specific matrix representation of the 
Lorentz operator.  In \cite{or98}, the Lorentz constraint
was solved by using a quite general matrix representation 
for the Bianchi IX cosmological model. 
The aim of this brief report is to improve an incorrect 
result obtained in \cite{mms98}
and to comment on the meaning and importance of the Lorentz 
constraint in $N=1$ supergravity. 

\section{Local Lorentz invariance}

The Lorentz constraint
\be
{\cal J}_{AB}\Psi[{\bf e}, {\bf \psi}] = 0 
\label{lorc}
\ee
simply states that the ground state of the wave functional must be invariant
with respect to local Lorentz transformations. Then, it is
clear that any Lorentz invariant quantity is a solution to
this constraint. The question is how to construct Lorentz
invariant wave functionals. The first thing one should notice 
is that the Lorentz operator ${\cal J}_{AB}$ acts
only on geometric objects which include vierbein Lorentz 
indices $A,B,...$. Then, any combinations of the
variables ${\bf e}$ and ${\bf \psi}$ in which all
the Lorentz indices are contracted represent  scalar
Lorentz invariants, i.e., solutions to the Lorentz
constraint. This strategy has been used to find explicit
wave functionals \cite{dho93,cs95} which also satisfy
the supersymmetric constraint. In general, it has been
argued that it is possible to construct the general 
solution of the Lorentz constraint by taking an expansion 
of the wave functional in bosonic and fermionic variables
in such a way that in all the terms the Lorentz indices
are contracted. However, this is obviously not true
because it would imply that only scalar quantities 
could be Lorentz invariants. If we would attempt to 
find all the Lorentz invariants which depend on
the field variables ${\bf e}$ and ${\bf \psi}$, we would
have to consider also invariant tensors of higher order and would
end up with an infinite number of possibilities, each one 
being a candidate to construct an expansion in 
bosonic and fermionic variables. We conclude that
almost any ``intelligent'' combination of bosonic and fermionic
variables is a Lorentz invariant. 
This is the reason why it is generally believed that
the Lorentz constraint in $N=1$ supergravity
is straightforward to satisfy \cite{cfop}.

On the other hand, if one insists in solving the Lorentz constraint explicitly, 
one first must find an appropriate representation for the Lorentz operator 
and the wave functional such that Eq.(\ref{lorc}) becomes plausible and
suitable to be solved. For instance, in \cite{or98} it was shown that the 
minimal matrix representation, which is in accordance with the algebra of the
gravitino field, leads to a wave functional that must be 
represented by a 64-component vector, satisfying 384 coupled algebraic equations.
Fortunately, this system of algebraic equations could be solved for the
Bianchi IX minisuperspace model, and generated only two non-vanishing independent 
components for the wave functional. Obviously, this method could lead to the
most general solution of the Lorentz constraint. In the relatively simple case
of the diagonal Bianchi IX model the resulting system of coupled algebraic equations
could be solved, but it is not clear whether less simple models would lead to 
a tractable set of equations. A different approach was used in \cite{mms98} 
which consists in choosing an $SO(3)$ basis for the vierbein ${\bf e}$ in which 
the Lorentz constraint contains only fermionic degrees of freedom \cite{pi78}
\begin{equation}
{\cal J}_{AB} = \frac{1}{2} {\phi_{[A}}^T \phi_{B]} 
\label{lc6}\, ,
\end{equation}
where $\phi_A$ are the densitizied local components of the fermionic field
$\phi_a = e\, e_a{}^\alpha \psi_\alpha$, with $e={}^{(3)}e= \det(e_a{}^{\alpha})$.
Here $a$ and $\alpha$ are spatial Lorentz and world indices, respectively. 
Moreover, since $\psi_0$ is a Lagrange multiplier [cf. Eq.(\ref{sgham})], one
should choose $\phi_0$ accordingly. Then from Eq.(\ref{lc6}) it seems natural 
to write the Lorentz constraint as \cite{mms98}
\begin{equation}
{\cal J}_{AB}\Psi[{\bf e},\psi] = \left( \begin{array}{cccc}
0&0&0&0\\
0&0&{\cal J}_{12}&{\cal J}_{13}\\
0&-{\cal J}_{12}&0&{\cal J}_{23}\\
0&-{\cal J}_{13}&- {\cal J}_{23}&0
\end{array}\right)
\left( \begin{array}{c}
 \Psi_I\\
 \Psi_{II}\\
 \Psi_{III}\\
 \Psi_{IV}\end{array}\right)= 0
\label{lc7}\, .
\end{equation}
Now the idea is to find a matrix representation for each of the components ${\cal J}_{ab}$ 
and $\Psi_I$, etc. 
such that the solution of of the Lorentz constraint (\ref{lc7}) 
is compatible with the particular choice of a basis which lead to Eq.(\ref{lc6}). 
Since in Eq.(\ref{lorg}) we used the product of $\gamma-$matrices to represent the
Lorentz generators it seems reasonable to use them again to represent each of the
non-vanishing components of the Lorentz operator ${\cal J}_{AB}$. To see if this is
possible we write explicitly the Lorentz constraint (\ref{lc7}) as
\begin{eqnarray}
{\cal J}_{12}\Psi_{III} &=& -{\cal J}_{13}\Psi_{IV}\label{sys1}\, ,\\
{\cal J}_{12}\Psi_{II} &=& {\cal J}_{23}\Psi_{IV}\label{sys2}\, , \\
{\cal J}_{13}\Psi_{II} &=& -{\cal J}_{23}\Psi_{III} \label{sys3}\, ,
\end{eqnarray}
and assuming that none of the components $\Psi_{II}$, $\Psi_{III}$, $\Psi_{IV}$
is zero, we obtain from Eqs.(\ref{sys1})-(\ref{sys3})
\be
{\cal J}_{13} ({\cal J}_{12})^{-1} {\cal J}_{23}=
{\cal J}_{23} ({\cal J}_{12})^{-1} {\cal J}_{13} \ .
\label{ntc}
\ee
This result is surprising because it implies a relationship between 
the components of the Lorentz operator, whereas the idea of the Lorentz
constraint is to impose conditions on the components of the wave functional.
However, if we could find a representation for the components of ${\cal J}_{ab}$
such that the condition (\ref{ntc}) becomes an identity, the contradiction would
be solved. In \cite{mms98} it was claimed that the choice 
\be
{\cal J}_{12} =   -\gamma^3 \gamma^0 \ , \ 
{\cal J}_{13} =  - \gamma^1 \gamma^3 \ , \ 
{\cal J}_{23} =  - \gamma^1 \gamma^0
\ee
is the solution to this problem.
However, a straightforward calculation shows that this choice is not 
a solution to the condition (\ref{ntc}), although it does lead to the 
choice 
\be
\phi_1 = -i\gamma^3 \, , \phi_2 =-i \gamma^1 \, , 
\phi_3 = -i \gamma^0
\ee
for the gravitino field which satisfies 
the definition (\ref{lc6}). By analyzing all possible products
of $\gamma-$matrices one can show that the choice 
\be
 {\cal J}_{12} =   -\gamma^1 \gamma^0 \ , \ 
{\cal J}_{13} =  - \gamma^1 \gamma^3 \ , \ 
{\cal J}_{23} =  - \gamma^2 \gamma^0
\ee
is the only solution (modulo permutations) to the condition (\ref{ntc}).
Unfortunately, this choice is not compatible with Eq.(\ref{lc6}), i.e.
there is no way to choose the spatial components of the gravitino field
$\phi_a$ proportional to a $\gamma-$matrix so that (\ref{lc6}) is 
satisfied (the index $a$ allows only three values whereas the last choice
for ${\cal J}_{ab}$ involves four different $\gamma-$matrices).  
Thus, we have shown that for non-vanishing components of the
wave functional the Lorentz constraint (\ref{lc7}) and the definition 
equation (\ref{lc6}) have no common solution which could be expressed 
in terms of $\gamma-$matrices. Unfortunately, 
this contradicts the results of \cite{mms98},
where the incorrect conclusions are due to a computational error. 

To solve this problem let us recall that
equation (\ref{lc6}) is valid only when the spatial part of the
local vierbein is an $SO(3)$ basis. This suggests that the 
spatial components ${\cal J}_{ab}$ should be related with the generators
of the rotation group. In fact, if we identify  the three
independent components of ${\cal J}_{ab}$ with the three generators $J^a$
of the ordinary rotation group by means of the relationship 
\be
{\cal J}_{ab} = i\varepsilon_{abc}J^c \ ,
\label{rotg}
\ee
with $ \varepsilon_{123}=1$, and use standard conventions of \cite{kaku}
\begin{eqnarray}
{\cal J}_{23} = iJ^1
 = \left( \begin{array}{cccc}
 0&0&0&0\\
0&0&0&0\\
0&0&0&1\\
0&0&-1&0
\end{array}\right)
\label{lo21}\, , \\ 
{\cal J}_{13}=-iJ^2  
= - \left( \begin{array}{cccc}
 0&0&0&0\\
0&0&0&-1\\
0&0&0&0\\
0&1&0&0
\end{array}\right)\label{lo22}\, , \\ 
{\cal J}_{12} = iJ^3
= \left( \begin{array}{cccc}
 0&0&0&0\\
0&0&1&0\\
0&-1&0&0\\
0&0&0&0
\end{array}\right)\label{lo23}\,, 
\end{eqnarray}
the Lorentz constraint, represented in Eqs.(\ref{sys1})--(\ref{sys3}),
can not be used to derive a condition similar to (\ref{ntc}), because
the representation (\ref{lo21})--(\ref{lo23}) is singular and consequently
the inverse  of the Lorentz operator is not defined.
Therefore, we must solve the Lorentz constraint (\ref{sys1})--(\ref{sys3})
explicitly as a system which imposes conditions on the components of the
wave functional. So this representation is now in agreement with the 
idea of the Lorentz constraint, as mentioned above. 

To find the solution of the Lorentz constraint in the above
representation we consider each of the components $\Psi_I$, $\Psi_{II}$,
$\Psi_{III}$, and $\Psi_{IV}$ as as a 4-component vector. Notice that
the component $\Psi_{I}$ is not affected by the action of the Lorentz 
constraint. For the remaining 12 components of $\Psi_{II}$,
$\Psi_{III}$, and $\Psi_{IV}$ we obtain a set of 9 algebraic equations
following from Eqs.(\ref{sys1})--(\ref{sys3}), namely
\begin{eqnarray}
\Psi_{III}^2 = \Psi_{IV}^2=0, \quad & &\Psi_{III}^3=-\Psi_{IV}^4 \ ,\\
\Psi_{II}^3 = \Psi_{IV}^3=0, \quad & & \Psi_{II}^2=-\Psi_{IV}^4 \ ,\\
\Psi_{II}^4 = \Psi_{III}^4=0, \quad & & \Psi_{II}^2=-\Psi_{III}^3 \ .
\end{eqnarray}
The right-hand side set of equations of this system 
implies that $\Psi_{II}^2=\Psi_{III}^3=\Psi_{IV}^4=0$, so that
each of these vectors possesses one non-vanishing component only. 
Then, the final form of the 
wave functional can be written as
\be
\Psi[{\bf e},\psi] =  
\left( \begin{array}{cccc}
\Psi_{I}^1 &\Psi_{II}^1 &\Psi_{III}^1 &\Psi_{IV}^1 \\
\Psi_{I}^2 &0&0&0\\
\Psi_{I}^3 &0&0&0\\
\Psi_{I}^4 &0&0&0
\end{array}\right)
\label{gsol} \ .
\ee
This represents the most general solution of the Lorentz constraint, when 
the spatial part of the local vierbein ${\bf e}$ corresponds to an $SO(3)$ 
basis. Notice that this final solution also explains why the condition 
(\ref{ntc}) for the components of the Lorentz operator is not valid.
In fact, it was obtained under the assumption that all the components
of $\Psi_{II}$, $\Psi_{III}$, and $\Psi_{IV}$ are non zero, a condition
which is satisfied by only 3 of these 12 components.

We see that this final form for $\Psi[{\bf e}, \psi]$ 
contains 7 different components which 
furthermore must satisfy the supersymmetric constraint.  
In a previous work \cite{msq} we showed that in the special case
of a midisuperspace described by a Gowdy cosmological model, the
number of independent components can be reduced to two, after
imposing the supersymmetric constraint. Incidentally, this final
number of independent components of the wave functional coincides
with the number obtained in \cite{or98} by applying a different
approach for the minisuperspace described by the Bianchi IX 
cosmological model.   

The general solution (\ref{gsol}) contains three 
quantities $\Psi_{II}^1$, $\Psi_{III}^1$, and $\Psi_{IV}^1$ which 
are scalars. This is in agreement with the intuitive idea mentioned above 
about the character of the possible solutions for the Lorentz constraint.
The four remaining degrees of freedom are contained in the 4-component
vector $(\Psi_I^1, \Psi_I^2,\Psi_I^3,\Psi_I^4)$ which is not affected 
by the Lorentz operator. The particular solution in which only one
degree of freedom remains $(\Psi_I^1, 0,0,0)$ and $\Psi_I^1$ is 
considered as a scalar, is called the ``rest-frame'' solution that 
has been intensively analyzed in the literature
because of its simplicity.

Finally, we must show that the singular representation 
(\ref{lo21})--(\ref{lo23}) is compatible with the definition 
of the components
of the Lorentz operator in terms of the components of the
gravitino field (\ref{lc6}). This condition turns out to be
identically satisfied. In fact, if we identify the gravitino
components in terms of the generators of rotation in the form
\be
\phi_a = 2i J^a \ ,
\label{phij}
\ee
introduce this equation into the right-hand side of Eq.(\ref{lc6}),
and use the relationship (\ref{rotg}) in the left-hand side 
of the same equation, we obtain [notice that $(J^a)^T = - J^a$]
\be
[J^a, J^b] = i\varepsilon_{abc}J^c \ .
\ee
This last expression is simply the  algebra of the generators
of the rotation group in the representation (\ref{lo21})--(\ref{lo23}).
This is an interesting result showing 
that the components of the Lorentz operator and the gravitino field 
are completely determined by the generators of the rotation group 
in such a way that the corresponding algebra represents the compatibility
condition for selecting  this specific representation.

\section{conclusions}
\label{con}

Although it is  very simple to find solutions to 
the Lorentz constraint of $N=1$ supergravity in terms of scalar quantities
constructed by means of bosonic and fermionic variables, one can try to solve
it explicitly in order to find more general solutions. For the approach 
in which one tries to represent the Lorentz operator and the gravitino field in 
terms of $\gamma-$matrices we showed that the solution presented in \cite{mms98} is 
not correct and that, in fact, there are no solutions at all. 
Instead, we propose to represent the Lorentz operator and the gravitino field 
in terms of the generators of the rotation group. In this case,  the solution is
simple as given in Eqs.(\ref{rotg}) and (\ref{phij}), and the compatibility 
condition (\ref{lc6}) coincides with the algebra satisfied by the generators
of the rotation group. 

For the present representation of the Lorentz operator to be valid it
is necessary that the spatial part of the local vierbein coincides with
an $SO(3)$ basis. In this case, the wave functional turns out to have at most
16 independent components. We solved the Lorentz constraint explicitly and
found that 9 components vanish so that the general solution contains 7 
degrees of freedom, four as the arbitrary components of a 4-vector and three
in the form of Lorentz invariant scalars.

\begin{acknowledgments}

We would like to thank O. Obreg\'on for helpful comments and literature hints.
This work was supported by CONACyT,  grants 42191--F and 36581--E.

\end{acknowledgments}

\end{document}